\documentclass[12pt]{article}
\usepackage[dvips]{graphicx}
\usepackage{pdproc}

  \makeatletter 
  \def\@cite#1{[#1]} 
  \makeatother    
\begin{document}

\renewcommand{\thefootnote}{\alph{footnote}}
\newcommand{\del}{\partial}
\newcommand{\Mp}{M_{\rm P}}
\newcommand{\Ms}{M_{\rm s}}

\title{
(Pseudo) Dirac neutrino masses in Supergravity }

\author{ ATHANASIOS DEDES}

\address{ Institute for Particle Physics Phenomenology (IPPP), \\
Durham DH1 3LE, United Kingdom
\\ {\rm E-mail: Athanasios.Dedes@durham.ac.uk}
}

\abstract{ In this talk, we discuss the idea of how Dirac or
Pseudo-Dirac neutrino masses arise naturally  with a correct
size, after the breaking of local N=1 supersymmetry.  }

\normalsize\baselineskip=15pt

\section{Introduction to the idea}

This talk is based on Ref.\cite{ADT} and on previous works done in
references therein.  In a nutshell,  we show  that Dirac
neutrino masses can arise from the K\"ahler potential of supergravity
and they are proportional to SUSY and  electroweak breaking scales.
Dirac neutrino masses of the correct size ($\le 0.05$ eV)
are obtained, provided that the ultraviolet cutoff, $M$, of the MSSM is
between the GUT and the heterotic string scale. The above is
guaranteed with the implimentation of an R-symmetry and the assumption
of Lepton number conservation. Breaking of Lepton number results in
Pseudo-Dirac neutrinos~\cite{Wolfenstein}. 

In order to obtain small Dirac neutrino masses
we use,  instead of the ``unnatural'' Dirac term 
\begin{equation}
W \supset L H_u N^c \;,\label{eq1}
\end{equation}
in the superpotential, the K\"ahler potential term
\begin{equation}
K \supset \frac{L H_u N^c}{M} + \frac{L H_d^* N^c}{M} + {\rm H.c} \;,
\end{equation}
where $M$ is the ultraviolet cutoff of the theory. After the
supersymmetry  and (radiative) electroweak symmetry breaking,  
we find Dirac neutrino masses with magnitude
\begin{equation}
m_\nu \simeq \frac{m_{3/2}}{M} \biggl 
[ \langle H_u \rangle + \langle H_d \rangle \biggr ] \sim 10^{-4}~{\rm eV} \;,
\label{eq3}
\end{equation}
with $M=M_{\rm P}=2.44 \times 10^{18}$ GeV, $\langle H_u \rangle =
m_{\rm top}=175$ GeV and $m_{3/2} \simeq 1$ TeV. The neutrino masses
obtained from Eq.(\ref{eq3}) are smaller than those implied from
atmosheric neutrino oscillation experiments, $m_\nu \simeq 0.05$ eV.
Turning the argument around, and using Eq.(\ref{eq3}), right size
neutrino masses imply a scale $M=5 \times 10^{15}$ GeV, just below the
GUT scale. However, with a more careful inspection,
in the context of full Supergravity we obtain
\begin{equation}
M=(4\times 10^{16} - 5 \times 10^{17})~{\rm GeV} \;,
\label{eq4}
\end{equation}
as explicitly shown below. The idea of obtaining neutrino masses
directly from the K\"ahler potential has been discussed in the
literature~\cite{Hall,Borzumati,Navarro,Kitano,Arnowitt,Smirnov,Russell} 
mainly within the context of global supersymmetry.
Dirac neutrino masses can also be derived 
in the  U(1) extended MSSM as described in  Ref.\cite{Ma}.

\section{Matter fermion mass terms in Supergravity}

To study the idea just sketched, we need to consider contributions to
fermion masses in a full supergravity set-up. To this end, consider
chiral superfields $S_i$ of some hidden sector which are responsible
for spontaneous breaking of Supergravity by acquiring v.e.v's, $S_i=M
\sigma_i$, and visible sector chiral superfields $y_\alpha$ of the
observable sector such as the leptons $L,E$, the Higgs $H_u,H_d$, and
the right handed neutrino $N^c$ superfield. 
%
The general formula for the fermion mass matrix $m_{\alpha\beta}$, for
an N=1 supersymmetric theory coupled to gravity, can be found in
standard textbooks~\cite{sugra,WB,Nilles}. Denoting the gravitino mass
as $m_{3/2} = \langle \frac{W^{\rm (h)}}{\Mp^2} \: {\rm exp}(K^{\rm
(h)}/2\Mp^2) \rangle$, and taking the flat limit $\Mp \to \infty$,
this formula reads
\begin{eqnarray}
m_{\alpha\beta}  &=&  \frac{1}{2}  \: \Biggl \{  
\frac{\del^2 W^{\rm (o)}}{\del y^a\del y^\beta}
- g^{\gamma\delta*}\frac{\del^3 K^{\rm (o)}}{\del y^\alpha\del 
y^\beta \del y^{\delta*}}
\frac{\del W^{\rm (o)}}{\del y^\gamma}
-  \frac{1}{M^2} \biggl [
g^{ij*} \frac{\del^3 K^{\rm (o)}}{\del y^\alpha \del y^\beta \del
\sigma^{j*}}\frac{\del W^{\rm (h)}}{\del \sigma^i}\biggl ]
 +\cdots \Biggr \} 
\nonumber \\[4mm]&&  \hspace{-1.5cm}  
+\,\, \frac{m_{3/2}}{2}\: \Biggl \{ 
\frac{\del^2 K^{\rm (o)}}{\del y^\alpha \del y^\beta} 
-g^{\gamma\delta*}\frac{\del^3 K^{\rm (o)}}{\del y^\alpha\del 
y^\beta \del y^{\delta*}}
\frac{\del K^{\rm (o)}}{\del y^\gamma}
-  \frac{1}{M^2} \biggl [
g^{ij*} \frac{\del^3 K^{\rm (o)}}{\del y^\alpha \del y^\beta \del
\sigma^{j*}}\frac{\del K^{\rm (h)}}{\del \sigma^i}\biggl ]
+ \cdots \Biggr \},
\label{master}
\end{eqnarray}
where the $``\cdots ``$ stand for terms involving the hidden-visible
mixed metric $g^{\gamma i*}$. In what follows we shall also assume 
that the metric $g^{ij*}=g^{\alpha\beta*}$ is diagonal.
Equation(\ref{master}) is our master formula. It is devided into two parts
: the first one ($1^{\rm st}$-line) exists in global supersymmetry while
the second one ($2^{\rm nd}$-line) is induced from the N=1 SUSY
coupled to gravity. The latter is proportional to the gravitino mass
$m_{3/2}$ and is the same as the first term with the replacement
$W^{(\rm h,o)} \to m_{3/2} K^{(\rm h,o)}$. The first term in first
line is the well known fermion superpotential mass term. If an
R-symmetry prohibits a bilinear term, it can be replaced by the first
term in the second line but multiplied by $m_{3/2}$. This is a well
known mechanism~\cite{GM,SW}.  A generalization of this mechanism is
employed below in deriving small (Pseudo) Dirac neutrino masses.

\section{(Pseudo) Dirac neutrino masses}

We first need  to prohibit the superpotential term in Eq.(\ref{eq1}).
We also want to generate the $\mu$-term with the same mechanism.
For simplicity, we assume that the observable superpotential
is a function only of observable fields, $W^{\rm (o)}(y)$. Then by 
imposing a discrete R-symmetry (see  Table.1 in~\cite{ADT}) we obtain
\begin{eqnarray}
W^{\rm (o)}(y) 
\ &\supset& \   
Y_E \: L H_d {E}^c + Y_D \: Q H_d {D}^c 
+ Y_U \: Q H_u {U}^c 
\;, 
\label{sp} \\[2mm]
K^{\rm (o)}(\sigma,\sigma^*,y,y^\dagger) \ &\supset& \  
 c_1(\sigma,\sigma^*) H_u H_d  
+\frac{c_2(\sigma,\sigma^*)}{M} L H_u {N}^c 
+ \frac{c_3(\sigma,\sigma^*)}{M}L H_d^* {N}^c
+ {\rm H.c} \;, \nonumber \\
\label{kp}
\end{eqnarray}
where $c_i(\sigma,\sigma^*)$ are functions of the hidden 
sector superfields.
Supergravity is spontaneously broken and, soon after that, electroweak
symmetry is radiatively broken. Use of the master formula,
Eq.(\ref{master}), results in the following Dirac neutrino masses
\begin{eqnarray}
m_\nu^{\rm D} \ &=& \ 
v \: \biggl (
\frac{m_{3/2} }{M}
\biggr )
\sin\beta \: \biggl [   c_2(\sigma,\sigma^*) - 
  c_1(\sigma,\sigma^*) c_3(\sigma,\sigma^*)
\biggr ] 
\nonumber \\[2mm] &-&  
 v \: \biggl  (
\frac{ F_S}{M^2}
\biggr )
\sin\beta \:\biggl [ 
\del_{\sigma^{\mbox{\tiny *}}}  c_2(\sigma,\sigma^*) 
+\cot\beta \: \del_{\sigma^{\mbox{\tiny *}}}  c_3(\sigma,\sigma^*) \biggr ] \;,
\label{md}
\end{eqnarray}
where $F_S=\del_S W^{\rm (h)} + m_{3/2}\, \del_S K^{\rm (h)}$.  In
local supersymmetry, for example, vanishing of the vacuum energy
implies that $F_S=\sqrt{3} \Mp m_{3/2}$. Right-size neutrino masses
imply that $M=(4\times 10^{16} - 5 \times 10^{17})~{\rm GeV}$, for
$100~{\rm GeV}\le m_{3/2}\le 10$ TeV. The second term in Eq.(\ref{md})
is enhanced by a factor $\Mp/M$ as compared to the first one. In
addition, the gravitino can be very light. Furthermore, the
non-holomorhic term proportional to $\del_{\sigma^*}c_3$ 
is dominant if $c_1=c_2=0$
or if cancelations take place. Another generic aspect (and maybe a problem)
of this mechanism is that soft breaking masses, as well as the
$\mu$-parameter, are expected to be of order $\tilde{m} \sim F_S/M \sim
10$ TeV, much larger than desired. If we relax the assumption of
lepton number conservation, then Dirac neutrinos obtained from the
K\"ahler potential can be ``polluted'' by the presence of Majorana
neutrino masses derived from extra non-renormalizable terms
\begin{eqnarray}
W^{\rm (o)}(y) 
\ &\supset& \   
 \frac{g_4}{M}(L H_u) (L H_u)  
\;, 
\label{sp2} \\[2mm]
K^{\rm (o)}(\sigma,\sigma^*,y,y^\dagger) \ &\supset& \  
\frac{c_4(\sigma,\sigma^*)}{M^3}W^{(h)}{N}^{c\: 2}
+ {\rm H.c.} \;
\label{kp2}
\end{eqnarray}
Let us assume assume that only the term (\ref{sp2}) is present. Then,
using the range for $M$ given above, we obtain, in addition to the
Dirac mass $m_\nu^{\rm D}\simeq 0.05~{\rm eV}$, a much smaller Majorana
mass $m_\nu^{\rm L} \simeq 10^{-5}$ eV. Assuming one generation, the
mass matrix reads
\begin{eqnarray}
\label{pseudomat}
\biggl ( \begin{array}{cc} m_\nu^{\rm L} & m_\nu^{\rm D} \\
                           m_\nu^{\rm D}             & 0 
\end{array} \biggr ) 
\;. \label{eq217}
\end{eqnarray}
Pseudo-Dirac neutrinos result, with mass difference $\delta m^2 \simeq
2 m_\nu^{\rm D} m_\nu^{\rm L} \simeq 10^{-6}~{\rm eV}^2$ and maximal
mixing. Astrophysical techniques to distinguish between Dirac and
Pseudo-Dirac neutrinos have been described in Ref.~\cite{Bell}.

\section{Questions/Conclusions}

The first question one asks is : what is the source of the
non-renormalizable operators i.e., the mechanism which generates the
scale M? The answer may be: a) radiative corrections to the K\"ahler
potential~\cite{Brignole}, b) effective operators in heterotic string
theories~\cite{Taylor}, or c) a GUT model. We have tried the last
possibility but we did not find an acceptable model that also complies
with the proton decay stability. Another important question is how to
account for non-trivial (maximal) neutrino mixing matrix.  The answer
to this question may be linked to the fact that the K\"ahler potential
parameters are not protected by the non-renormalization theorem, and
vertex corrections may induce large flavour mixing through RGE
running. Another question concerning leptogenesis with Dirac neutrinos
has been (and is currently being) addressed in Ref.\cite{Lindner}.
Furthermore, a question concerning ``fundamentals'' is that, in order
to put the light pseudo-Dirac neutrino mechanism to work and for the
superpotential and K\"ahler potential to obtain the particular form
given in Eqs.({\ref{sp},\ref{kp}), one has to impose a symmetry such
as the $R$-symmetry in Table~1 of Ref.\cite{ADT}.
Is this symmetry broken and, if so, where and how?  One
attractive answer is to gauge a $U(1)$ anomaly-free $R$-symmetry. It
is known \cite{Dreiner} that such a symmetry must be broken at scales
$M \le \Mp$, which favours our case, with no effects from gauging the
$R$-symmetry remaining at low energies.

The bottom-line of the idea presented in Ref.~\cite{ADT}
is that neutrino masses are not filtered through unknown Majorana mass
terms but carry direct information about the structure of the K\"ahler
metric (more accurately of the Christoffel symbols of the
metric). (Pseudo) Dirac neutrino with mass of the correct size can
naturally arise in supergravity.

\bibliographystyle{plain}

\end{document}